\documentclass{aa}
\usepackage{graphicx}
\usepackage{subfigure}
\usepackage{amssymb}
\usepackage{amsmath}
\usepackage{natbib}
\bibpunct{(}{)}{;}{a}{}{,}   

\newcommand{\eqb}{\begin{eqnarray}}
\newcommand{\eqe}{\end{eqnarray}}

\newcommand{\melec}{m_{\rm e}}
\newcommand{\mprot}{m_{\rm p}}

\newcommand{\gammamax}{\gamma_{\rm max}}
\newcommand{\gammamin}{\gamma_{\rm min}}
\newcommand{\tgammamin}{\tilde{\gamma}_{\rm min}}
\newcommand{\tgammamax}{\tilde{\gamma}_{\rm max}}
\newcommand{\emin}{\epsilon_{\rm min}}
\newcommand{\emax}{\epsilon_{\rm max}}

\newcommand{\eB}{\epsilon_{\rm B}}
\newcommand{\ee}{\epsilon_{\rm e}}

\newcommand{\Rdec}{R_{\rm d}}

\begin{document}

\title{Effects of a low electron distribution cutoff on multiwavelength spectra and light curves of GRB afterglows}
\author{M. Petropoulou$^1$, A. Mastichiadis$^1$ \& T. Piran$^2$
} \institute{$\phantom{a}^1$ Department of Physics, University of
Athens,
Panepistimiopolis, GR 15783 Zografos, Greece\\
$\phantom{a}^2$ The Racah Institute of Physics, The Hebrew
University, Jerusalem, Israel}
 \offprints{A. Mastichiadis }
\date{Received ... / Accepted ...}
\abstract{}{ We investigate the behavior of the frequency-centered
light curves expected within the standard model of Gamma Ray Bursts
allowing the maximum electron energy ($\gammamax$) to be a free
parameter permitted to take low values.}{We solve the spatially
averaged kinetic equations which describe the simultaneous evolution
of particles and photons, obtaining the multi-wavelength spectra as
a function of time. From these we construct the frequency-centered
light curves giving emphasis in the X-ray and optical bands. }{We
show that in cases where $\gammamax$ takes low values, the produced
X-ray light curves show a plateau as the synchrotron component gives
its place to the Synhro Self-Compton one in the X-ray band.}{}

\titlerunning{Effects of a low electron distribution cutoff on GRB afterglows}
\authorrunning{Petropoulou, Mastichiadis \& Piran}
\maketitle

\section{Introduction}

Gamma-Ray Bursts (GRBs) are attributed to a release of a very large
amount of energy ($\sim 10^{51}-10^{52}$ ergs) into a small region
of space ($\lesssim 100$ km) over a short period of time ($\sim
10-10^2$ s for long GRBs and $\lesssim 2$ s for short GRBs). These
energetic events have two characteristic radiative signatures: (i)
the prompt $\gamma$-ray and (ii) the afterglow emission. The
detection of high energy photons ($\epsilon_{\gamma} \gg 1$ MeV)
implies sources of radiation moving at relativistic speeds with
Lorentz factors $\Gamma$ exceeding 100 \citep{fenimore93,
lithwick01,Piran99}.

While many issues concerning the prompt emission are still open, the
afterglow, i.e. the lower energy long lasting emission, is believed
to arise from the interaction of the relativistic ejecta with the
ambient matter and can be adequately described by the so-called
`standard' model \citep{rees_mesz92, paczynski_rhoads93,
mesz_rees97}. According to this, the relativistic blast wave
produced from the explosion can energize the external medium, i.e.
accelerate electrons (and possibly protons) to high energies and
generate magnetic fields. The relativistic electrons radiate by
synchrotron and inverse Compton radiation which is essentially the
observed afterglow emission. However, in order to calculate the
radiated photon spectra, one needs a detailed prescription of the
electron distribution function and of the magnetic field. This is
usually done by postulating that the electrons have a power law
distribution between a minimum ($\gammamin$) and a maximum
($\gammamax$) cutoff with an overall energy density content which is
a a fixed fraction (usually denoted by $\ee$) of the total
post-shock internal energy density $U$
while an analogous argument can be made for the magnetic field
energy which takes a fraction $\eB$ of $U$. A significant amount of
work has been performed by many researchers in calculating the
multiwavelength spectra and light curves of GRB afterglows either
based directly on the above prescription
\citep{dermchiang98, sariPiran98,
panai_mesz98,wijers_galama99, dermbottcher00, panai_kumar00,
granotsari02} or using different variations \citep{granot_kumar06,
fan_piran06, panaietal06, zhang06, nousek06}.

In the present paper we focus on the effects that a low $\gammamax$
will bring on the multiwavelength spectra and light curves of the
afterglow emission. This has not been treated thus far as it is
implicitly assumed that $\gammamax$ is very large and its radiative
signature does not contribute to any observable band. However, if it
has a low value, then a break might appear successively in various
energy bands of diminishing energy as the synchrotron component
gives gradually its position to the SSC one. This will produce light
curves which are not any more pure power laws but have more
complicated shapes.

The paper is structured as follows. Is \S2 we describe the
principles of the model and discuss, in a qualitative way, some of
the results.
 In \S3 we quantify the above
and we derive some analytical relations between the initial
parameters, which, when satisfied, will produce different types of
X-ray light curves. In \S4 we make a tentative connection of our
results to observations. Finally in \S5 we conclude and give a brief
discussion of the basic points of the present work.


\section{The Model}
\label{model}

\subsection{First principles}

The general framework of the model we present here is based on the
standard GRB afterglow model, albeit with some modifications
regarding mainly the approach to the physical problem (
\cite{fanetal08, me09} -- henceforth PM09). For the sake of
completeness we repeat here its basic premises: as the Relativistic
Blast Wave (RBW) usually associated with GRB afterglows is
expanding, it accelerates by some unspecified mechanism electrons of
the circumstellar medium to high energies. These are assumed to be
injected behind the shock front in a volume of radius R containing a
tangled magnetic field B. The particles suffer radiative and
adiabatic losses, evolving with radius. At the same time they emit
radiation by synchrotron and inverse Compton radiation. Therefore,
at each radius there is a coupling between electrons and photons, in
the sense that the electron distribution function determines the
photon spectrum and, at the same time, the photons determine the
electron distribution function through inverse Compton losses and,
possibly, pair reinjection. The usual procedure of approaching the
problem is to solve simultaneously two coupled kinetic equations for
the distribution functions of electrons and photons which, when
solved, give the aforementioned quantities as functions of radius
and energy. The most relevant physical processes which are included
are: electron synchrotron radiation,  synchrotron self absorption,
inverse Compton scattering (both in the Thomson and Klein-Nishina
regimes), photon-photon pair production and  adiabatic losses (for a
more detailed discussion regarding the physical processes see
\cite{mastikirk95, peer04}).

To obtain the multiwavelength (hereafter MW) spectrum of GRBs at
each radius $r$ of the relativistic blast wave one needs to specify
the Lorentz factor of the flow $\Gamma(r)$, the comoving radius of
the source $R=r/\Gamma$, the magnetic field strength $B(r)$
 -- determined indirectly
through the parameter $\eB$, and three parameters related to the
electron injection, i.e. their total power -- determined by the
parameter $\ee$, their slope $p$ of the power law at injection and
the maximum cutoff of their distribution $\gammamax$ -- the minimum
cutoff $\gammamin$ is defined in terms of
the other  parameters by equation (\ref{gminfull}). 

\subsection{Multiwavelength spectra}

\begin{figure}[h]
 \centering
\includegraphics[width=8.5cm,height=5.9cm]{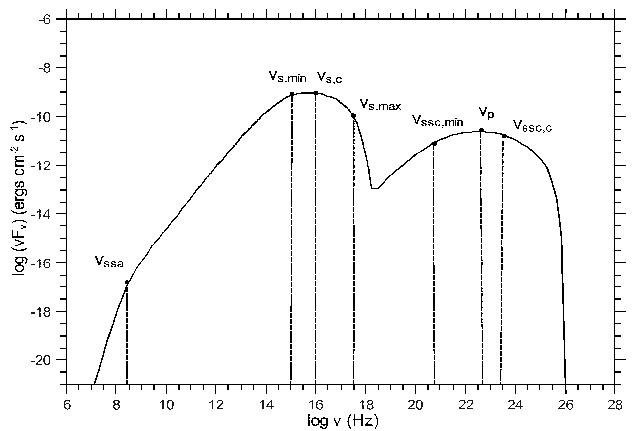}
 \caption{
Multiwavelength spectrum expected for a case where the upper
electron cutoff is not much greater than the lower one -- in the
present case $\gammamax=10^4$ and $\gamma_{\rm min,0}=5.6 \times
10^3$. The complete set of the parameters used is: $E_0=10^{54}\ \rm
ergs$, $\Gamma_0=400$, $n_0=1 \ \rm part/cm^3$, $\eB=0.005$,
$\ee=0.01$ and $p=2.3$. Both synchrotron and SSC components of the
spectrum appear to be continuously curved without any clear power
law segments (at least for frequencies below $\nu_{\rm s, min}$ and
$\nu_{\rm ssc,min}$ respectively) mainly due to the proximity of
$\gammamax$ and $\gammamin$. The characteristic frequencies
indicated in the figure are discussed in the text.} \label{vFv}
\end{figure}

The approach described above allows one to calculate the photon MW
spectra self-consistently while it inherently addresses questions
about whether the electrons are in the fast or slow cooling regime.
Moreover it can calculate the spectrum in various regimes without
resorting in a piecewise succession of broken power laws which
becomes problematic in cases where the characteristic frequencies
are rather close to one another. Finally, it takes into account SSC
losses which, as PM09 have shown -- see also \cite{sariesin01}, can
alter significantly the electron spectrum and therefore, the photon
spectrum, even in the well-studied synchrotron regime.

Some of the above are exemplified in Fig.~\ref{vFv} which depicts a
generic case of a MW spectrum in the case of a power-law electron
injection with $\gammamax$ not much greater than $\gammamin$. As the
parameters have been chosen so as the radiating electrons to be
partly in the uncooled regime, the synchrotron component consists,
at least in theory \citep{sariPiran98}, of four segments: $(1)$
synchrotron self-absorbed part below $\nu_{\rm ssa}$, $(2)$ typical
synchrotron low energy part for $\nu_{\rm ssa}  < \nu < \nu_{\rm s,
min}$, $(3)$ uncooled part for $\nu_{\rm s, min} < \nu < \nu_{\rm s,
c}$ and $(4)$ cooled part for $\nu_{\rm s, c} < \nu < \nu_{\rm s,
max}$. One could add here (5) the synchrotron exponential cutoff
which appears above $\nu_{\rm s, max}$. Between these segments power
laws of different slopes should appear, however due to the proximity
of the lower and upper electron cutoffs, the turnovers in frequency
have been smeared out and the spectrum appears to curve continuously
without any clearly defined power-law regime, at least for
frequencies above $\nu_{\rm s, min}$.

The same holds for the SSC component: this is much broader, as
expected, than the synchrotron one  and it also shows a continuous
curvature. In Fig.~\ref{vFv} three characteristic frequencies of the
SSC component are indicated: $\nu_{\rm ssc,
min}=\frac{4}{3}\gammamin^2 \nu_{\rm s, min}$, $\nu_{\rm ssc,
c}=\frac{4}{3}\gamma_{\rm c}^2\nu_{\rm s, c}$ and the peak frequency
$\nu_{\rm p}$. Sari \& Esin (2001) have shown analytically that the
peak frequency of the SSC component in the slow cooling regime is
given by $\nu_{\rm ssc, c}$. However, this is not expected when a
relatively low $\gammamax$ is taken into account, as in the case of
Fig.~\ref{vFv}. An analytic calculation (which can be found in
detail in the Appendix) of the SSC peak frequency can also be done,
in the case of an electron pure power law distribution, having
limits
 between $\gammamin$ and $\gammamax$. The
SSC peak frequency is then given by $\frac{4}{3}\gammamax^2 \nu_{\rm
s, min}$ which equals to $1.4 \times 10^{23}$ Hz for the example
given in Fig.~\ref{vFv}. The numerically calculated peak frequency
is however different: $\nu_{\rm p}=4.4 \times 10^{22} \ \rm Hz
\approx \frac{4}{3} \gamma_{\rm c}^2\nu_{\rm s, min}$. This simple
example shows that the presence of the Compton logarithm
\citep{gould79} combined with the fact that the electron
distribution might have at least two breaks, one at $\gammamin$ and
one at $\gamma_{\rm c}$, makes a simple analytic approach
complicated. Nevertheless, for the specific example one can clearly
see in $\nu F_\nu$ units a rising part, a broad peak and a declining
part.


\subsection{Light curves}

Based on the above it is straightforward to construct light curves
at various frequencies. However, it is worth describing
qualitatively a case where cutoff effects in the synchrotron
component appear due to a low value of $\gammamax$. Assume a certain
observing window between two frequencies. Assume also that the
initial parameters are such as initially the flux in this window to
be dominated by the synchrotron component. As time evolves in the
observer's frame, he/she will first observe the various parts of the
synchrotron component passing through it, as first was suggested by
\citet{sariPiran98}. Therefore at some point in time, which we will
call $t_{\rm br, 1}$, the combination of $\Gamma$, $B$ and
$\gammamax$ will be such that the flux in the observing window will
be dominated by the exponential cutoff of the synchrotron component.
This will result in a natural steepening of the light curve. In the
hypothetical case where the SSC component is absent, the observer
would have seen an ever increasing steepening of both the spectral
index and of the light curve until the flux would drop to very low
levels, below the sensitivity of any instrument. In reality,
however, at some level the SSC component must appear in the
observing window and start dominating the flux there. This would
naturally result in a flattening of the light curve as the observer
starts sampling photons not from the synchrotron but from the SSC
component. Depending on the relation between $\gammamin$ and
$\gammamax$,  at time $t=t_{\rm br, 1}$ the SSC component could be,
broadly speaking, in any of its three spectral regimes (i.e. rising,
peak or declining). As we elaborate in the next section, if the SSC
component becomes dominant while still early in its rising part, the
light curve will show a sharp turnover which will result in a
shallow decline of the flux, i.e a plateau. If, on the other hand,
the transition from the synchrotron to SSC occurs while the latter
is close to the peak or during its declining part, the light curve
will show a more gentle flattening.

\begin{figure}[h]
 \centering
\includegraphics[width=0.45\textwidth]{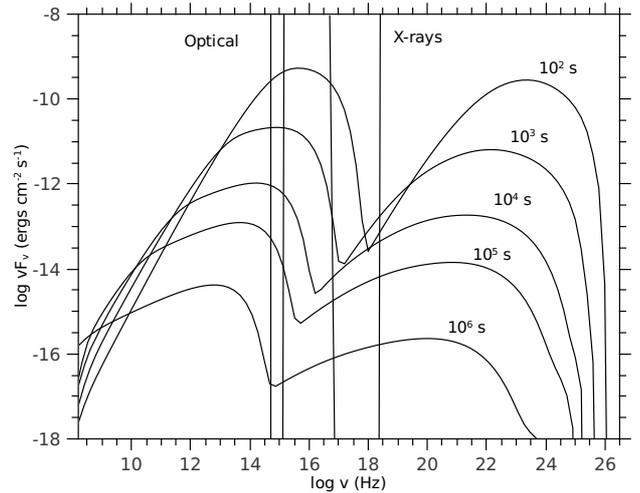}
\caption{Multiwavelength spectra at observer times from $10^2$ s to
$10^6$ s (top to bottom). For the parameters used see
text. X-ray and optical windows that correspond to the observing
energy ranges of XRT $(0.3-10)$ keV and UVOT $(170-650)$ nm
respectively, are also shown.}
 \label{MWspec}
\end{figure}

\begin{figure}[h]
 \centering
\includegraphics[width=0.45\textwidth]{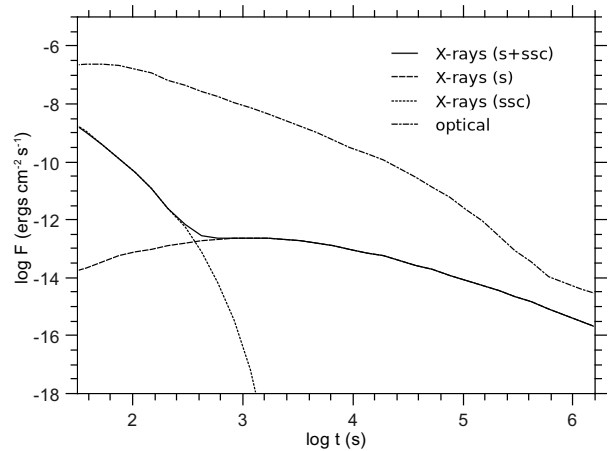}
\caption{X-ray
(solid line) and optical (dashed-dotted line) light curve
corresponding to the same case as in Fig.\ref{MWspec}. The
contribution of the synchrotron (dashed line) and SSC (dotted line)
components to the total X-ray flux are also shown. The optical light
curve is shifted by +2.5 units in logarithm for reasons of better
display.}
 \label{XUVlc}
\end{figure}

Figures~\ref{MWspec} and \ref{XUVlc} present the results of a run
showing an X-ray plateau. Figure~\ref{MWspec} shows snapshots of the
MW spectra at five different times, obtained for a GRB at $z=1$ with
$E_0=10^{54}$ ergs, $\Gamma_0=400$, $n_0=1$ part/$\rm cm^3$,
$\eB=0.001$, $\ee=0.025$, $p=2.3$ and $\gammamax=2.24 \times 10^4$.
One can see that at early times the X-ray band is dominated by the
tail of the synchrotron component which in about $t\sim 10^3$ s has
given its position entirely to the SSC one. This trend is repeated
once more at much later times ($t \gtrsim 5 \times 10^5$ s) in the
optical band. Figure~\ref{XUVlc} shows the corresponding optical and
X-ray light curves. The latter shows a clear plateau which gradually
steepens due to the concavity of the SSC component. On the other
hand, the optical light curve shows the `standard' unbroken
power-law behavior until $t\sim 5\times 10^5$ s and flattens off
slightly at even later times, as the SSC component takes over.

The above results show that in order the X-ray light curves to show
plateaus, at least within the context of our model, the SSC
component should 
be rather flat in the X-ray energy range at the time the synchrotron
component decays due to its exponential cutoff. Since the above
conditions implicate the magnetic field $B(r)$, and the lower and
upper cutoffs of the electron injection, $\gammamin$ and $\gammamax$
respectively, one can quantify the above conditions using the
parameters of the standard afterglow model. We proceed to do this in
the next section.

\section{Effects of $\gammamax$}
The upper limit of the electron distribution $\gammamax$ has not
been so far considered in GRB afterglow models as a dynamic
parameter. In this section we will show how a relatively low
$\gammamax$ can affect the MW spectra and the corresponding X-ray
and optical light curves.
We assume that the afterglow is produced by an adiabatic
relativistic blast wave decelerating while interacting with the ISM.
During this phase the evolution of the RBW is described by the
self-similar solution of \cite{blandMckee76}:
 \eqb \Gamma(r) = \Gamma_0 \left(\frac{r}{R_d}
\right)^{-3/2} \quad \textrm{if} \ r>R_d,\eqe where
 \eqb
 R_d = \left(\frac{3E_0}{4\pi n_0 m_p c^2
\Gamma_{0}^{2}}\right)^{1/3}. \eqe
 Using the relation
\eqb \int_{t_0}^t \rm dt=\int_{r_0}^{r} \frac{\rm
 dr'}{2c\Gamma^2(r')}
\eqe
 and neglecting the term $t_{\rm d}=\frac{\Rdec}{2c\Gamma_0^2}$ compared
to $\int_{\Rdec}^r \frac{\rm dr'}{2c\Gamma^2(r')}$ we find that \eqb
 r \approx  (8ct \Gamma_{0}^{2}
R_{d}^{3})^{1/4}.
 \eqe
For the strength of the magnetic field (in the fluid frame), we
adopt the usual form
 \eqb
 B(r)=\sqrt{32 \pi n_0 m_p \eB} c \Gamma(r).
 \eqe
 The observed synchrotron frequency corresponding to $\gammamax$
 is given by:
\eqb \nu_{\rm s,\rm max} &=& \Gamma \frac{eB}{2 \pi m_e c}
\gammamax^2  \nonumber \\
&=& \left(\frac{3E_0 n_0 m_p \eB^{2} e^4}{32 \pi^3 c^5
m_{e}^{4}}\right)^{1/4} \gammamax^2 t^{-3/4}.\eqe
 We consider an X-ray window that corresponds
to the observing energy range of XRT, i.e $(0.3-10)$ keV. We are
interested in the time $t^{(\rm x)}_{\rm coff,\rm s}$ when photons
belonging to the synchrotron exponential cutoff which forms above
$\nu_{\rm s,\rm max}$ (see Fig.~1) cross a characteristic frequency
of the X-ray band, say $\nu_{\rm x} \approx 6.3 \times 10^{17}$ Hz.
So we set $\nu_{\rm s,coff} = A \nu_{\rm s,\rm max}$ where $A$ is a
numerical factor of order $5-10$ which determines how deep into the
exponential cutoff the particular synchrotron photons are. Then the
expression for $t^{(\rm x)}_{\rm coff,\rm s}$ becomes
 \eqb
\label{ts}
 t^{(\rm x)}_{\rm coff,\rm s} &\approx& \left(\frac{3E_0 n_0 m_p
\eB^{2} e^4}{32 \pi^3 c^5 m_{e}^{4}}\right)^{1/3} \nu_{\rm x}^{-4/3}
A^{4/3} \gammamax^{8/3} \cdot
\label{txcoff} \eqe
Demanding that the characteristic time $ t^{(\rm x)}_{\rm coff,\rm
s}$ lies in a time interval of the general form: \eqb t_k \leq
t^{(\rm x)}_{\rm coff,\rm s} \leq t_{k+1}, \label{early} \eqe
 where $t_{k}=10^k$ s, equations (\ref{ts}) and
 (\ref{early}) combine to give the first constraining relation for $\gammamax$:
 \eqb
 \label{time}
  g_1(E_0,n_0,\eB) \leq \gammamax \leq
 g_2(E_0,n_0,\eB),
 \eqe
 where $g_{1,2}(E_0,n_0,\eB)= C_{1,2} \ 10^{3k/8} \ A^{-1/2} E_{0,54}^{-1/8} n_{0,0}^{-1/8}
 \epsilon_{\rm B,-2}^{-1/4}$ with $C_1=4.1\times 10^3$ and $C_2=9.7\times
 10^3$. One should keep in mind that relation (\ref{early}) is valid
only for values of the real variable $k$ which ensure that time
$t_k$ is larger than the deceleration time $t_{\rm d}$. Here and
throughout the text, the convention $Q_{x}\equiv Q/10^x$
 has been adopted in cgs units. It is worth noting that for typical
 values of $E_{0,54}=n_{0,0}=\epsilon_{\rm B,-2}=1$ and $k=2$ the
 maximum electron Lorentz factor lies between $8.1\times 10^3$ and $1.9 \times
 10^4$, which is relatively low.
 Although the above double inequality involves three free parameters of the model, the
dependence on two of them, i.e on $E_0$ and $n_0$ is very weak.

 The SSC component peaks at a characteristic frequency which
depends on the Compton logarithm first introduced by
\citet{gould79}. In the simple case where the electron distribution
is given by a pure power law between $\gammamin$ and $\gammamax$ and
the scatterings happen in the Thomson regime, it is found that the
peak frequency of the SSC component is $\nu_{\rm p
}=\frac{4}{3}\gammamax^2\nu_{\rm s, \rm min}$ (see Appendix for a
detailed calculation). In a generic case where the electron
distribution shows a cooling break, the calculation of the SSC peak
frequency is more complicated (see discussion in \S2.2). However, in
all cases the minimum frequency of the main SSC branch, i.e
$\nu_{\rm ssc,\rm min}=\frac{4}{3} \gammamin^2 \nu_{\rm s,\rm min}$
is below (slow cooling regime) or at least equal (fast cooling
regime) to the peak frequency. For that reason we choose to quantify
the lower energy part of the SSC component as $b \nu_{\rm ssc,\rm
min}$, where $b$ is a numerical factor of order $0.001-0.01$. In
order to proceed we need to use an expression for $\gammamin$. So
far in GRB afterglow models, $\gammamax$ was not treated as a
`dynamic' parameter in the sense, that its signature would not be
observed in the X-ray energy range and below, as it was taken to be
much larger than $\gammamin$. For that reason the approximate
expression for $\gammamin$ \citep{sariPiran98}
 \eqb
\label{gminapprox}
 \gammamin^{\rm approx} = \ee
 \frac{\mprot}{\melec}\frac{p-2}{p-1}\Gamma(r)
 \eqe was safely used. However, in our work where
we also examine cases with $\gammamax$ only a few times greater than
$\gammamin$, in the numerical code we use the accurate expression
which is solution to the equation: \eqb \label{gminfull}
\frac{\gammamin^{2-p}-\gammamax^{2-p}}{\gammamin^{1-p}-\gammamax^{1-p}}
= \ee f_p \frac{\mprot}{\melec} \Gamma(r), \eqe where
$f_p=\frac{p-2}{p-1}$.
As the solution to the equation above has no explicit analytical form, we
proceed first to find a constraint of $\gammamax$ using the
approximate analytical expression given by eq. (\ref{gminapprox}). 
For the low energy part of the SSC component we can now write:
 \eqb
 b\nu_{\rm ssc,\rm min} & \approx & \frac{2eb}{3 \pi m_e}\sqrt{32 \pi
n_0 m_p \eB} \left(\ee
\frac{m_p}{m_e}f_p\right)^4 \cdot{} \nonumber \\
& & {}\cdot  \left(\frac{3E_0}{2048 \pi n_0 m_p c^5}\right)^{3/4}
t^{-9/4}.\eqe
 In order to have a flattening of the light curve after
the first break, this low energy part of the SSC component must
appear in the X-ray band about the time when the synchrotron one
decays. In complete analogy to $t^{(\rm x)}_{\rm coff,\rm s}$, the
observed time $t^{(\rm x)}_{\rm ssc}$ at which the frequency $b
\nu_{\rm ssc,\rm min}$ enters the X-ray band is determined by the
relation
 \eqb t^{(\rm x)}_{\rm ssc}&=& \nu_{\rm
x}^{-4/9} \left(\frac{2eb}{3 \pi m_e}\right)^{4/9}(32 \pi n_0 m_p
\eB)^{2/9} \left(\ee
\frac{m_p}{m_e} f_p\right)^{16/9}\cdot {}\nonumber \\
& & {} \left(\frac{3E_0}{2048 \pi n_0 m_p c^5}\right)^{1/3}.\eqe

 The ratio of the two characteristic times is given
by: \eqb
 \frac{t^{(\rm x)}_{\rm ssc}}{t^{(\rm x)}_{\rm coff,\rm s}}&=&\left[\left(\nu_{\rm x}
\frac{m_p}{m_e}\right)^2 \frac{\pi m_p}{6 e^2}\right]^{4/9} \!\!\!\!
b^{4/9} A^{-4/3}\gammamax^{-8/3}\left(\frac{f_p^{4} \ee^{4}}{n_0
\eB}\right)^{4/9} \nonumber \\ \\
 &\approx& 2 \times 10^{16} \
b^{4/9} A^{-4/3}\gammamax^{-8/3}\left(\frac{f_p^{4} \ee^{4}}{n_0
\eB}\right)^{4/9}.\eqe
 When the two
timescales $t^{(\rm x)}_{\rm coff, \rm s}$ and $t^{(\rm x)}_{\rm
ssc}$ are of the same order, a break will appear in the X-ray light
curve, as the synchrotron component gives its place to the low
energy part of the SSC one. Allowing
 \eqb 0.2 < \frac{t^{(\rm x)}_{\rm ssc}}{t^{(\rm x)}_{\rm coff,\rm s}} < 1.0\eqe
we obtain one more constrain for $\gammamax$, i.e.
 \eqb
\label{tratio} h_1(n_0,\ee,\eB,f_p) < \gammamax <
h_2(n_0,\ee,\eB,f_p), \label{timeratio}
 \eqe
 where the functions $h_{1,2}$ are defined as $h_{1,2} = K_{1,2} \ A^{-1/2} b^{1/6} f_p^{2/3} \epsilon_{\rm e,-1}^{2/3}
 n_{0,0}^{-1/6} \epsilon_{\rm B,-2}^{-1/6}$ with $K_1=6 \times 10^5$ and $K_2=1.1 \times 10^6$.
 The above relations make the tacit assumption that the scatterings
occur in the Thomson regime, at least for the lower energy part of
the SSC spectrum. 
For that reason, we impose one more constraint on $\gammamax$
demanding that at least for $t \sim t^{(\rm x)}_{\rm coff, \rm s}$
the assumption of scattering in the Thomson regime is valid. For
that we use the dimensionless factor
 \eqb x &=& \frac{\gammamin h \nu'_{\rm
s,\rm min}}{m_e c^2} \nonumber
\\ &=& \frac{\hbar e }{8 m_{e}^{2} c^2}\left(\frac{3 E_0
\eB}{c^5}\right)^{1/2} \left(f_p \frac{\ee
 m_p}{m_e}\right)^3 t^{-3/2}, \eqe
where primed quantities are measured in the comoving frame.
 At $t=t^{(\rm x)}_{\rm coff,\rm s}$ the $x$ parameter is given
by:
 \eqb x &=& \frac{\hbar f_p^{3} m_{p}^{3}}{8 e m_{e}^{3} c^2}\left(\frac{32
\pi^3}{ m_p}\right)^{1/2} \nu_{\rm x}^{2}\  \ee^{3} A^{-2}
\gammamax^{-4}(n_0 \eB)^{-1/2}.
\eqe
Thus, when $x < 1$ we find that: \eqb
 \label{thomson}
\gammamax \gtrsim 50 \ A^{-1/2} n_{0,0}^{-1/8} f_p^{3/4}
\epsilon_{\rm e,-1}^{3/4} \epsilon_{\rm B,-2}^{-1/8}.
 \eqe
The above relation for typical parameter values poses only a weak
constraint on $\gammamax$. The same holds even if the accurate
expression for $\gammamin$ was used. For that reason from here on
 we will not take into consideration the
constraining relation (\ref{thomson}).

\begin{figure}
\centering
\includegraphics[height=5cm, width=7cm]{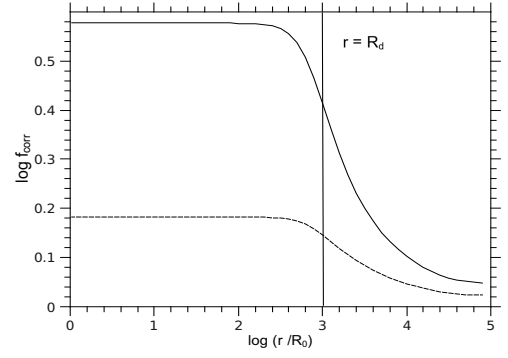}
\caption{Log-Log plot of $f_{\rm mod}$ as a function of radius $r$ (in units of the
initial radius $R_0=10^{14}$ cm) for two
values of the ratio $\frac{\gammamax}{\gamma_{\rm min,0}}$: 1.32 (solid line) and 33 (dashed
line). Other parameters used are: $E_0=10^{54}$ ergs, $\Gamma_0=400$, $n_0=1$ part/$\rm cm^3$,
$\eB=0.001$, $\ee=0.01$ and $p=2.3$. The deceleration radius is also shown. }
\label{fcorr}
\end{figure}

Having determined the constraining relations for $\gammamax$ using
the approximate form of $\gammamin$, we can now estimate the
corrections introduced,
 after taking into account the
accurate expression of it. For the purposes of our analytic analysis
we model it as:
 \eqb
 \label{gminacc}
 \gammamin=f_{\rm mod}\gammamin^{\rm approx},
 \eqe
 where the term $f_{\rm mod}$ is a function of the radius $r$ and of the
 ratio $\frac{\gammamax}{\gamma_{\rm min,0}}$; $\gamma_{\rm min,0}$ is
 the initial minimum Lorentz factor
 of the electrons. Function $f_{\rm mod}$ is plotted, for illustrative reasons,
  against radius $r$ for two
 values of $\frac{\gammamax}{\gamma_{\rm min,0}}$ in Fig.~\ref{fcorr}. It is evident that
 $f_{\rm mod} \rightarrow 1$ in the limit
 of $\gammamax >> \gamma_{\rm min, 0}$, as expected.
As our analysis holds for the deceleration phase of the RBW, we can
simplify the calculations further by averaging the function $f_{\rm
mod}$ over $\tilde{r}=\log r$ for $r>R_d$: \eqb \bar{f}_{\rm mod}
\left(\frac{\gammamax}{\gamma_{\rm min,0}}\right)
=\frac{\int_{\substack {r>R_d}} d\tilde{r} f_{\rm mod}\left(
\tilde{r},\frac{\gammamax}{\gamma_{\rm min,0}}\right)}{\int_{r>R_d}
d\tilde{r}}\cdot \eqe This average modifying parameter can be used
in eq. (\ref{gminacc}) instead of $f_{\rm mod}$. Thus, constraining
relation (\ref{tratio}) becomes: \eqb \label{tratio2}
 \bar{h}_1<\gammamax<\bar{h}_2,
\eqe where $\bar{h}_{1,2}=\bar{f}^{\phantom{a} 2/3}_{\rm mod}
h_{1,2}$. Table~\ref{table1} shows some indicative values of the
correction introduced in relation (\ref{tratio2}) for $E_0=10^{54}$
ergs, $\Gamma_0=400$, $n_0=1$ part/$\rm cm^3$, $\eB=0.001$
$\ee=0.01$ and $p=2.3$.
\begin{table}[h!]
\newcommand\T{\rule{0pt}{2.6ex}}
\newcommand\B{\rule[-1.2ex]{0pt}{0pt}}
\centering
\begin{tabular}{l l l}
\hline
$\log \gammamax$ \T \B & $\log\left(\gammamax/\gamma_{\rm min,0}\right)$ & $\frac{2}{3}\log \bar{f}_{\rm mod}$\\
\hline
4.15 \T & 0.495& 0.087\\
4.35 & 0.787 & 0.072\\
5.35 \B & 2.00 & 0.030\\
\hline
\end{tabular}
\caption{Indicative values of the correction introduced in the constraining relation (\ref{tratio2}).
For the parameters used see text.}
\label{table1}
\end{table}

\begin{figure}[h]
 \centering
 \includegraphics[width=0.45\textwidth]{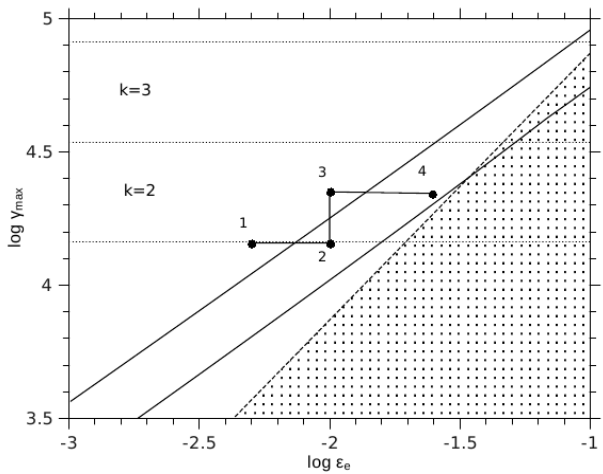}
 \caption{Initial value parameter space of
$\gammamax$ and $\epsilon_e$ for $E_0=10^{54}$ ergs, $\Gamma_0=400,
n_0 = 1 \ \rm part/cm^3, \epsilon_B=0.001$ and $p=2.3$.
 The numerical factors chosen here are $A=8, b=0.001$ (see text for their definition).
 Dotted lines correspond to the constraining relation (\ref{time}), solid lines to
  (\ref{tratio2}) and the dashed line
 sets the boundary of the striped area, which is not permitted as it
 it leads to $\gammamax=\gamma_{\rm min,0}$. The horizontal zones labeled by $k=2$ and
$k=3$ correspond to transition times $t^{(\rm x)}_{\rm coff,s}$
lying in the time intervals $10^2-10^3$ s and $10^3-10^4$s
respectively. The `coordinates' $(\ee,\gammamax)$ of the points 1 to
4 marked on the plot are: $\{(0.005,1.41 \times 10^4),\ (0.01,
1.41\times 10^4),\ (0.01, 2.24 \times10^4),\ (0.025,
2.24\times10^4)\}$ respectively . The corresponding X-ray and
optical light curves of the aforementioned points are presented in
the following Figs.~\ref{lcX} and \ref{lcV}.}
 \label{par_space}
\end{figure}

Using eq. (\ref{time}) and (\ref{tratio2})
 we can plot $\gammamax$ versus $\ee$ for fixed values of $E_0, n_0, p$ and
$\eB$, creating a parameter space shown in Fig.~\ref{par_space}.
 The other parameters used are the same as those used in Table~\ref{table1} above.
The curves defined by eq. (\ref{time}) and (\ref{tratio2})
create distinctive areas on the parameter space. For initial values chosen from the striped area,
eq. (\ref{gminfull}) has as solution $\gamma_{\rm min,0}=\gammamax$ with
no physical meaning. In that sense the striped area
is not permitted.
The horizontal zones labeled by $k=2$ and $k=3$ are related to cases
where the transition time $t^{(\rm x)}_{\rm coff,s}$ lies between
$10^2-10^3$ s and $10^3-10^4$ s respectively. A choice of $k<2$
would 
correspond to breaks occuring at even earlier times. We call a
`plateau strip' the tilted zone, as a choice of a pair $(\ee, \gammamax)$ from it leads to X-ray
light curves which show a shallow decline phase, i.e a `plateau'.
Moreover, if this pair is chosen from the intersection of the `plateau strip' with a horizontal
zone labeled by $k=2$ for example, the plateau phase will begin at a time between $10^2-10^3$ s and so on.
Finally, a choice of
pairs from the area outside the tilted zone leads to X-ray
light curves with 
a single change of slope.

\begin{figure}[h]
 \centering
\includegraphics [width=0.45\textwidth]{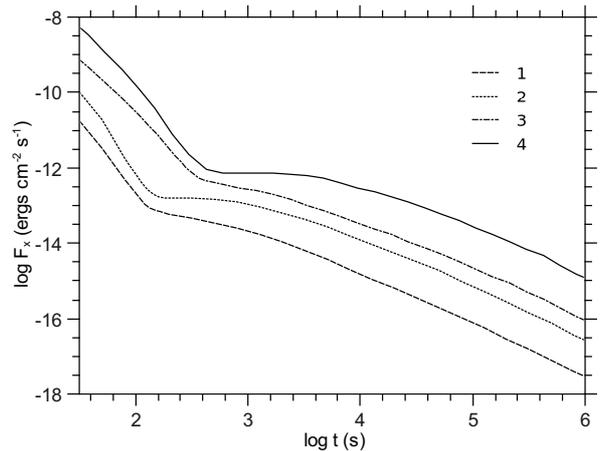}
\caption{X-ray
light curves for different set of parameters corresponding to the
points from 1 to 4 (bottom to top) marked on panel (b) of
Fig.~\ref{par_space}. The first break of the X-ray light curves 1
and 2 occurs at around $125$ s, while for the light curves 3 and 4
occurs later at $t\sim 650$ s. For clarity reasons light curves $1$,
$3$ and $4$ are plotted with an offset of $-0.2$, $+0.4$ and $+0.5$
in logarithmic units of flux respectively.} \label{lcX}
\end{figure}

For a better inspection of the parameter space, we have chosen pairs
from four distinctive regions of the diagram and show the
corresponding X-ray and optical light curves in Figs.~\ref{lcX} and
~\ref{lcV}. As we move from point 1 to 2, i.e by increasing $\ee$
while keeping $\gammamax$ constant, the break times (see
Fig.~\ref{lcX}) remain constant while the light curve shapes exhibit
a transition as the decay becomes flatter leading to a plateau
phase. From point 2 to 3, the break is shifted towards later times
while the light curve becomes, once again steep. This is to be as
expected since point 3 lies outside the `plateau strip'. Finally, as
we move from point 3 to 4 the slope of the light curve after the
break becomes flat and a plateau is produced. This behavior is the
same with the already discussed transition from point 1 to 2, with
the sole exception that now the break occurs at later times. Note
that for all the numerical runs presented in the present work, we
use the accurate expression for $\gammamin$ by solving eq.
(\ref{gminfull}). For that we have included in our numerical code a
subroutine that utilizes a combination of the bisection and
Newton-Raphson Method \citep{pressetal}.

\begin{figure}[h!]
\centering
\includegraphics[width=8.8cm, height=7.2cm]{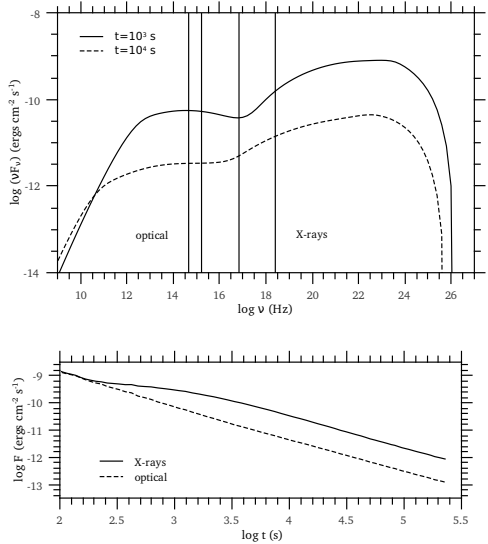}
\caption{Multiwavelength spectra at observer times $10^3$ s and
$10^4$ s (top) and light curves for the X-ray and optical band
(bottom) for an SSC dominated afterglow. For the parameters used see
text. The X-ray and optical windows are as given in
Fig.~\ref{MWspec}.}\label{highX}
\end{figure}

\begin{figure}[h]
 \centering
\includegraphics [width=0.45\textwidth]{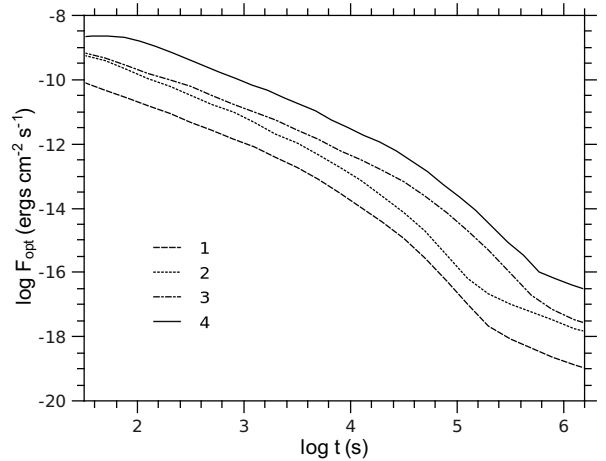}
\caption{Optical light curves for different set of parameters
corresponding to the points 1 to 4 (bottom to top) marked on panel (b) of
Fig.~\ref{par_space}. For clarity reasons light curves $1$, $3$ and
$4$ are plotted with an offset of $-0.4$, $+0.2$ and $+0.5$ in
logarithmic units of flux respectively.}
 \label{lcV}
\end{figure}

One could also make some comments on the effects of the other free parameters, which were
assumed fixed in Fig.~\ref{par_space}, will have on the parameter
space. For example, a possible change of $E_0$ will affect only the
dotted lines in Fig. ~\ref{par_space}. The effect will not be of
great importance as the inequality (\ref{time}) has a very weak
dependance on $E_0$. In general, an increase of $\eB$ will shift the
`plateau strip' downwards. Although the parameter $\Gamma_0$ does
not appear explicitly in the constraining relations (\ref{time}) and
(\ref{tratio2}) has an effect on the relative positions of the
striped area and the `plateau
 strip'.
 Finally, the external number density $n_0$
may not affect severely the appearance of the parameter space, but
has an important effect on the ratio of the synchrotron to the
 SSC flux. We note that as it was shown in PM09, an increase of
 $n_0$ makes the afterglow more Compton dominated. This possibility
has also been discussed by \cite{panai_kumar00} and \cite{sariesin01}.
 A denser external medium, increases in $\nu F_\nu$ units the X-ray flux over the
optical. Such high values of external density are required by our model
 in cases where the X-ray light curve shows a shallow phase and the 
ratio $L_{X}/L_{\rm opt} >1$.
Figure~\ref{highX} shows such a case. The parameters used are
$E_0=10^{54}$ ergs, $\Gamma_0=100$, $n_0=10^3$, $\eB=10^{-5}$,
$\ee=0.032$ and $\gammamax=8.2\times10^4$. For $t>10^3$ s, when the
spectral eivolution in the X-ray energy band is not very important,
the optical flux lies approximately one order of magnitude below the
X-ray one.

As the RBW decelerates the synchrotron and SSC emission from the
forward shock becomes weaker and softer (see Fig.~\ref{MWspec}). As
a result, the synchrotron cutoff will enter at some instant $t_{\rm
coff,s}^{(\rm opt)}$ in the optical band. It is straightforward to
show, under the assumptions of the problem, i.e. when $B\propto
r^{-3/2}$ and $\gammamax$ independent of $r$, that this is related
to $t_{\rm coff,s}^{(\rm x)}$ by:
 \eqb t_{\rm coff,s}^{(\rm
opt)}& =& \left(\frac{\nu_{\rm x}}{\nu_{\rm opt}}\right)^{4/3}t_{\rm
coff,s}^{(\rm x)} \eqe or, equivalently, by \eqb
 t_{\rm coff,s}^{(\rm opt)}& \approx &
2\times10^3 \ t_{\rm coff,s}^{(\rm x)}, \label{topt}
 \eqe
where a typical optical frequency $\nu_{\rm opt} = 2\times 10^{15}$
Hz was used. For the cases represented by points 1 and 2 in
Fig.~\ref{par_space} one finds $t_{\rm coff,s}^{(\rm x)}\approx 125$
s (see also Fig.~\ref{lcX}). Thus, a break in the optical light
curve is expected at around $2.5\times 10^5$ s, which is confirmed
numerically and it is shown in Fig.~\ref{lcV}. The other two optical
light curves in the same figure show a break at correspondingly
later times ($\gtrsim 6.3 \times 10^5$ s). We should also note that
the break of the optical light curve will not neccessarily be of a
`plateau' type. This is due to the fact that the emerging SSC
component has become steeper since the time it has emerged in the
X-ray window (see Fig.~2 for the time evolution of the SSC
component). A natural outcome of our model is that the optical light
curves do not exhibit a break during the plateau phase of the
corresponding one in X-rays, as it was first noted by
\cite{fan_piran06}. Evidence for chromatic breaks in X-ray light curves were also
discussed in \cite{panaietal06}.

Chromatic breaks in the optical
and X-ray light curves can be produced by our model with an expected
time difference given by eq. (\ref{topt}). As we discuss in the last
section, this is a strong constrain which reflects the validity of
the model's assumptions.

\section{Basic results}

\begin{figure}[h!]
\centering
\includegraphics[width=0.45\textwidth]{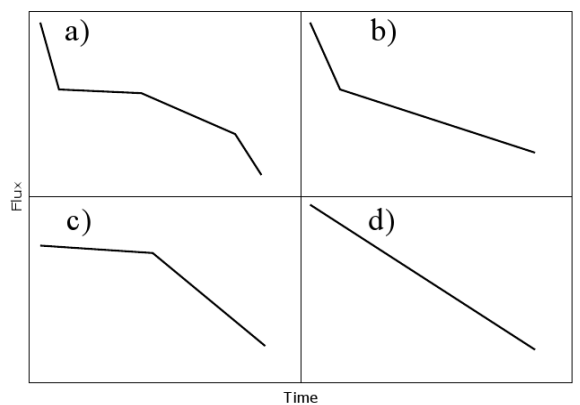}
\caption{Schematic diagram from \cite{Evans09} showing the different light
curve morphologies observed.}
\label{sximatika}
\end{figure}

\begin{figure}[h!]
\centering
\includegraphics[width=9cm, height=8cm]{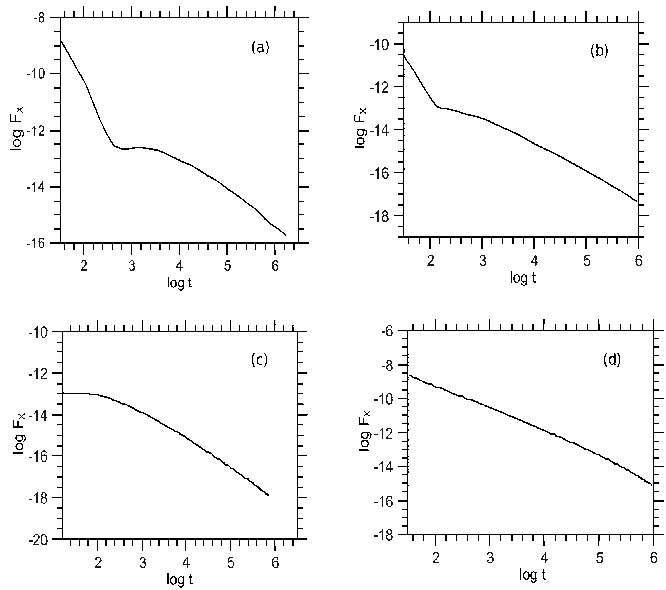}
\caption{Different light curve morphologies obtained using
our numerical code corresponding to points from different regions of the parameter
space shown in Fig.~\ref{par_space}. Light curves from each panel can be tentatively
compared to the corresponding ones of Fig.~\ref{sximatika}. For the parameters used see text.}
\label{sximatika2}
\end{figure}

It is well known that the X-ray and optical light curves of GRB
afterglows show a wide range of behaviors. Especially in the X-ray
regime \citep{nousek06, Evans09} the behavior can be quite
perplexing (see Fig.~\ref{sximatika}) with many cases showing a
plateau which is quite difficult to interpret
within the context of the standard model. 
 The analysis performed in \S3 can tentatively be of some relevance to
these observations. Using, for instance the example set by
Fig.~\ref{par_space} as a guide, we can say that plateaus appear as
long as the values of $\gammamax$ and $\ee$ are chosen from the
region defined by the two tilted lines. Moving inside this region
from the lower left to the upper right, plateaus appear at
progressively longer times. Furthermore, a choice of the initial
parameters outside this region (for example, from the left top
corner of Fig.~\ref{par_space}) leads to afterglows without a
plateau phase. Fig.~\ref{sximatika2} shows different types of X-ray
light curves obtained using our numerical code corresponding to
points from different regions of the parameter space of
Fig.~\ref{par_space}. Light curves in panels (a) and (b) correspond
to points (4) and (1) already
 shown in Fig.~\ref{par_space}, while light curves of panels (c) and (d) are
obtained using $(\ee,\gammamax)=(0.0032, 6.3\times10^3)$ and $(0.01, 10^6)$
 respectively.\footnote{All other parameters used are the same as in Fig.~\ref{par_space} except
 for the case presented in panel (c) where $\Gamma_0=800$ was used.}
  A tentative comparison of our model light curves
 to those of Fig.~\ref{sximatika} can be made.

\begin{figure}[!]
\centering
\includegraphics[width=8.5cm, height=8cm]{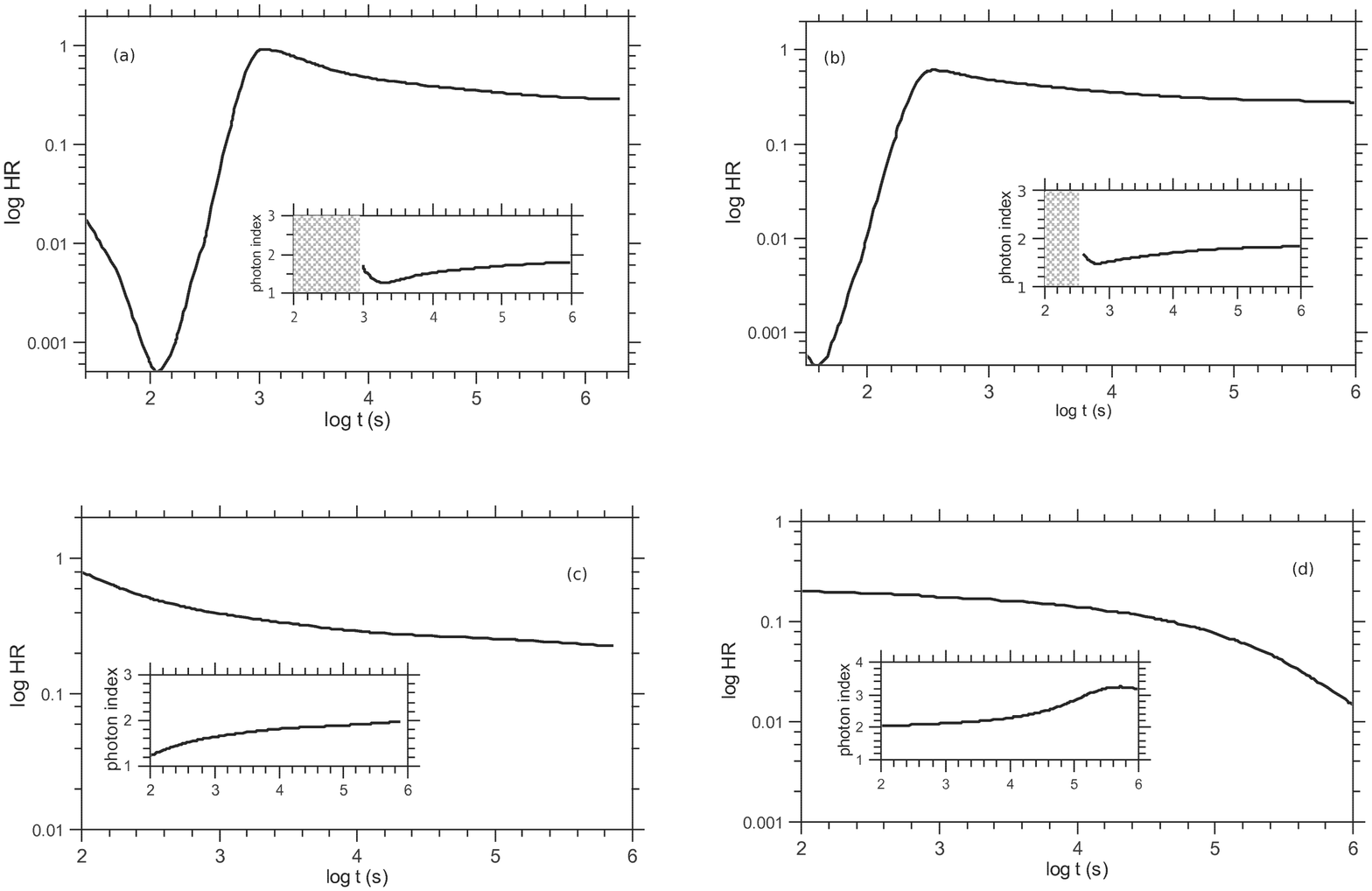}
 \caption{Time evolution of the hardness ratio for each of the cases presented
in Fig.~\ref{sximatika2}. For the definition of the hardness ratio used see text. The insert
 in each panel shows the evolution of the corresponding photon index. 
The shaded areas in the inserts imply that during this period our model spectra cannot be approximated by a single power law and 
therefore a specific photon index could not be attributed to the spectrum.}
 \label{HR}
\end{figure}

As the model presented here produces multiwavelength spectra at each
instant, we can use it to calculate the evolution of the expected
X-ray hardness ratio defined as the ratio of counts in the 1.5-10
keV to the counts in the 0.1-1.5 keV band \citep{Evans09,evans10}.
Figure~\ref{HR} shows the time evolution of the hardness ratio
for each of the example cases shown in Fig.~\ref{sximatika2}. 
Time evolution of the corresponding photon index 
is also shown in the inserts of Fig.~\ref{HR}, whenever the spectral
shape allows its viable calculation at the particular time (for a more detailed discussion
on the shape of our X-ray model spectra see \S2). 

For X-ray light curves with a distinctive `plateau' phase, we find that the
spectral evolution shows a characteristic trend as the X-ray window
is first dominated by the synchrotron and later by the SSC
component. This can be seen in panel (a) of
 Fig.~\ref{HR}. At very early times both the soft and
hard X-ray bands are dominated by the synchrotron photons -- however
the hard band is affected first by the synchrotron cutoff and this
has as a result the decrease of the hardness ratio. During this phase
the spectrum in X-rays is shaped by an exponential cutoff (synchrotron emission) and an emerging
flat power law component (SSC emission). Thus, it cannot be simply approximated by a single 
power law and `labeled' by a photon index (shaded area in the insert).
At later times, the SSC component starts appearing in the hard band while the
decreasing synchrotron component dominates the soft one, resulting
in an increase of the hardness ratio. Finally, at even later times
both bands are dominated by the SSC component, whose low energy part
can be approximated by a flat power law, and due to its gradual
steepening the hardness ratio appears to decrease gently.

In cases where the X-ray flux decays as a power law with time, as in panel (d) of Fig.~\ref{sximatika2}, we
find no significant spectral evolution. The photon index is approximately constant almost for three or four
decades in time, as the power law segment of the synchrotron component
 dominates until late times in the X-rays (see insert in panel (d) of Fig.~\ref{HR}).
The other two cases presented in panels (b) and (c) of Fig.~\ref{HR} lie somewhere inbetween
the two aforrementioned example cases.

Although the qualitative evolution
of the photon index with time is a robust feature of our model
, its specific value depends on the value of the other model parameters, such as
the slope of the electron energy spectrum. In all our runs we have used a typical value of
$p=2.3$.

\section{Summary/Discussion}

In the present paper we have investigated the role that the upper
cutoff of the electron injection can play in the evolution of the
multiwavelength spectra and light curves of GRB afterglows. 
For this we have solved self-consistently the kinetic equations
that govern  the electron
evolution and photon radiation as a function of distance
(see also \cite{fanetal08}
and PM09). This  approach can address successfully the
effects of the electron cutoff radiation on the light curves.

We have
shown that depending on the adopted value of $\gammamax$
the X-ray lightcurves can show one of the following behaviours:

\begin{enumerate}
\item
In cases where $\gammamax$ is not much greater than the
lower cutoff $\gammamin$, the X-ray light curves show three distinct
phases. First a fast drop phase which corresponds to the exponential
cutoff of the synchrotron component. Then a plateau phase which is
caused from the gradual dominance of the SSC component over the
decaying synchrotron. Finally, a more gradual power-law decay which
corresponds to the normal evolution of the SSC component.
The analytical approach used in \S3 and summarized in
Fig.~\ref{par_space} shows that, depending on the initial
parameters,  there might be a narrow strip in phase space that
produces X-ray plateaus. In all cases we found that $\gammamax$
should  be no more than a factor of 10 larger than $\gammamin$.\\
\item 
If $\gammamax$ does not satisfy the above condition 
but it is still close to the `plateau strip', then the X-ray afterglows
do not show a plateau but simply a change in the power law decay,
i.e. the three phases degenerate into two.\\
\item 
In cases where
$\gammamax\gg\gammamin$ we find that the X-ray afterglows will be
dominated until very late times by the synchrotron component, i.e.
we obtain the standard afterglow picture.
\end{enumerate}

These trends are exemplified in Figures~\ref{lcX} and \ref{sximatika2}. As far
as the flux decay in the optical regime is concerned, the present model predicts that
the optical light curve will mimic the X-ray one with two major differences (see
Fig.~\ref{lcV}). First, the break in the optical light curve will come at much later 
times (see eq.(\ref{topt})) and second, after the break, the optical light curve will
not neccessarily show a plateau. Note however, that both of these statements are
based on the assumption that $\gammamax$ remains constant throughout the RBW evolution.

In this respect, $\gammamax$ emerges as one of the important parameters
of the afterglow evolution as its choice can control critically the
behaviour of the X-ray light curves.

The evolution of the X-ray hardness ratio and spectral indices
were presented in Fig~\ref{HR}. 
Our model derived hardness ratio shows a characteristic signature which is
compatible with observations at least during the early stages
\citep{butler2007, liang07}. On the other hand it is still
inconclusive regarding the late stages (P. O'Brien -- private
communication), as the transition from the plateau to the normal
decay shows an evolution in the hardness ratio by a factor of two,
which is not the case for several individual GRBs (e.g.
\cite{vaughan06, liang07}). However, as preliminary calculations
 have shown (Petropoulou, Mastichiadis $\&$ Piran (2011) -- to appear
 in the proceedings of the $25^{\rm th}$ Texas Symposium held last December
 in Heidelberg), this is greatly relaxed in the case where $\gammamax$ is
allowed to increase with radius. As the study of such cases is
beyond the scope of the present paper, we will treat this in a
forthcoming publication.

 It is important to emphasize at this point that the close
relation we found between X-ray light curves exhibiting a plateau
phase and a $\gammamax$ not much greater than $\gammamin$, is what
actually differentiates our model from other works (e.g. long
lasting energy injection into the forward shock - refreshed shock
models \citep{zhang06}, late prompt emission
\citep{ghisellini07,ghisellini08}, geometrical effects
\citep{eichler06, granot_konigl06}, non-standard deceleration of the
bulk Lorentz factor due to the Compton-drag force
\citep{mastichiadiskazanas09}, dust scattering \citep{shao07} and
others).
Some recent PIC simulations show that most of the particles
are accelerated into a relativistic Maxwellian, while a small
fraction of them is injected to a power law high energy tail, whose
high energy cutoff is approximately only one order of magnitude
larger than the low one \citep{spitkovsky08,spitkovsky09}. However,
this is not a final result, as the present numerical simulations
have not reached yet a steady state, where $\gammamax$ is expected
to be larger than the one found so far. It is interesting to note
also that when we set $\gammamax$ not much larger than $\gammamin$
the qualitative behavior of our results is similar to the one
obtained when one replaces the power law injection with a
relativistic Maxwellian \citep{giannios09}. All other parameters
used in our work are the same as in the standard GRB afterglow
model.

Spectral evolution during the steep early phase, which is
observed in a significant number of GRB afterglows (see
\cite{zhangliang07} for a systematic study of 44 steeply decaying
X-ray afterglows), is an inevitable outcome of our model, as in our
present work the early steep decay of the X-ray light curve is
atrributed to the emission from the external shock. Moreover there
is some observational evidence for a possible smooth connection of
the early afterglow to the prompt emission \citep{Barthelmy05}. In
such a case the present work could be seen as an extension of the
supercritical model \citep{mastichiadiskazanas09} to the afterglow
regime. Alternatively it can result from a gradual transition towards the 
end of the prompt phase to an external shock emission. Evidence
for such a transition is seen in several examples where the
 extrapolated BAT light curve is not connected
 to early-time XRT light curve \citep{tagliaferri05}.
This fact combined with a strong spectral evolution at early times
suggests that the two emissions are produced by either different
mechanisms or in different regions. Such a transition from the
prompt to the afterglow emission could be explained within the
context of the internal/external shock scenario (\cite{Piran99,
piran05}-- see also \cite{dermer07} for a discussion on the possible
scenarios).

Finally we would like to present potentially problematic 
points of our model -- some of these could be addressed
with future observations and further analysis.
\begin{enumerate}
\item 
In the case of an afterglow with a plateau phase we find 
that the X-ray spectrum cannot be represented by a simple power-law
during the steep decay-early plateau phase. The spectral shape
at this stage can more accurately be described by a steep component
coming from the synchrotron cutoff plus a flat power-law
coming from the emerging SSC component -- see Fig.~\ref{MWspec} at early times. \\
\item
Some preliminary efforts in fitting lightcurves and hardness ratios
\footnote{Although a successfully fit of the HR could mot neccessarily mean
a succesful spectral fit, we preferred to model the former than the
latter as in many cases our X-ray spectra cannot be reproduced well by
a single power-law -- see also the previous point.}
have shown that, in some cases, our model can successfully reproduce both
(e.g. GRB 060512 -- see Petropoulou et al. 2011). However, in other cases
the model can fit successfully only the lightcurve while the HR 
fit is poorer (e.g. GRB 050713B).\\
\item
If the light curve shows an abrupt break (either a plateau phase or 
a single change in the slope decay)
 and $L_X/L_{\rm opt}>1$, then our model
requires high values of the external density ($n\ge 100~\rm{cm}^{-3}$).
This could be problematic in cases of GRB afterglows
 for which low values of $N_H$ are derived \citep{schady07}. We note however that,
in all other cases the above constrain of our model is relaxed.\\
\item
For those GRBs where the early steep decay is smoothly connected 
to the prompt emission phase, the model requires that the 
late prompt emission is already dominated by an
 external shock emission. This arises naturally in the external
shock model or it requires a transition, during the late prompt phase
, from in the internal to external shock.
\end{enumerate}

Concluding we could say that the consideration of the electron
distribution's upper energy cutoff as another free parameter in the standard
afterglow model brings many interesting features in the light curve / spectral
behavior of GRB afterglows. All these are related to the evolution of the
relativistic electron distribution and eventually point out at the acceleration
mechanism at work. We plan to deal with this issue in a forthcoming publication.

\begin{acknowledgements}
AM would like to thank TP for hospitality during a visit in Hebrew
University. We thank Drs.  P. Evans, P. O'Brien
 and A. Zezas for useful
discussions and Dr. D. Giannios for comments on the manuscript. This
work made use of data supplied by the UK Swift Science Data Centre
at the University of Leicester. This research has been co-financed by the
European Union (European Social Fund - ESF) and Greek national funds through
the operational Program `Education and Lifelong Learning' of NSRF - Research 
Funding Program: Heracleitus II. 

\end{acknowledgements}

\appendix
\section{Synchrotron Self-Compton spectrum of a power-law electron distribution}
 We assume a spherically symmetric source of radius $R$
with a magnetic field $B$ randomly oriented. Synchrotron radiation
is produced by an isotropic electron distribution, of the form \eqb
 n_{\rm e}=k_{\rm e}\gamma^{-p} \qquad \gammamin<\gamma<\gammamax
 \eqe
 where $p$ is the exponent of the power-law spectrum, $\gammamin, \gammamax$ are the cutoffs of the distribution
  and $k_{\rm e}$ is a normalization factor that determines the
  electron density and depends on radius $r$ in the case of an inhomogeneous source. The synchrotron photon production rate per
 unit energy and per unit volume is then given by
 \eqb
 \label{jsyn}
 \frac{dN_{\rm s}}{d\epsilon dt}= \frac{2 e^3}{m_{\rm e} h^2 c^2}
 \left(\frac{3e}{4\pi m_{\rm e}c}\right)^{\frac{p-1}{2}}
  \!\!\alpha(p)\ k_{\rm e}B^{\frac{p+1}{2}}\epsilon^{-\frac{p+1}{2}}
 \eqe
 where $\alpha(p)$ is a combination of $\Gamma$ - functions (see eq.~(4.60) of \citet{blumgould70}). Equation (\ref{jsyn}) holds
for photon energies not close to the low- and
 high- energy ends of the spectrum - $\epsilon_s \gammamin^2$ and $\epsilon_s \gammamax^2$ respectively, where $\epsilon_s=\frac{eB}{2 \pi m_{\rm e}c}$.
 The number density of
 synchrotron photons in the source is given by:
 \eqb
n_{\rm s}(\epsilon,r)=t_R\frac{dN_{\rm s}}{d\epsilon
dt}=\tilde{n}_{\rm s}(r)\epsilon^{-\frac{p+1}{2}}
 \eqe
 where $t_R$ is the crossing time of the source and $\tilde{n}_{\rm
 s}$ is the energy independent factor on the right hand side of
 eq.~(\ref{jsyn}). The dependence on radius comes from the quantities
 $k_e$ and $B$.
The explicit functional form of the photon density can be found in
\citet{gould79}. However, for the calculation of the Compton
synchrotron logarithm we can safely proceed using
 the spatial averaged photon density
 \eqb
 \bar{n}_{\rm s}(\epsilon)=<\!\!\tilde{n}_{\rm s}\!\!>\epsilon^{-\frac{p+1}{2}}
 \eqe
 where $<\!\!\tilde{n}_{\rm s}\!\!>=\frac{3}{R^3}\int{dr \ r^2 \tilde{n}_{\rm
 s}(r)}$. We assume further that the Inverse Compton emissivity is given by a
$\delta$ - function centered at the mean energy of an upscattered
 synchrotron photon of energy $\epsilon$
\eqb j_{\rm ic}(\epsilon_1; \gamma, \epsilon)=A \epsilon_1
\delta(\epsilon_1-4/3\gamma^2 \epsilon)
 \eqe
where $A$ is a normalization factor. Then the total SSC power per
unit energy emitted is found by
 \eqb
 \label{ICtot}
 J_{\rm ic}(\epsilon_1)= \int d\epsilon \ n_{\rm
 s}(\epsilon,r)\int d\gamma \ N_{\rm e}(\gamma)j_{\rm ic}(\epsilon_1; \gamma,\epsilon)
 \eqe
In the above equation $N_{\rm e}=K_{\rm e}\gamma^{-p}$ is the total
number of electrons per Lorentz factor $\gamma$ in the source. The
normalization factor $K_{\rm e}$ is related to $k_{\rm e}$ through
the integral $4 \pi \int dr \ r^2 k_{\rm e}(r)$. In order to
simplify further the calculation of the integral in (\ref{ICtot}) we
use the average photon density
 \eqb
 \label{ICtot2}
J_{\rm ic}(\epsilon_1)=A K_{\rm e} <\!\!\tilde{n}_{\rm s}\!\!>
 \int_{\emin}^{\emax}{d \epsilon
\epsilon^{-\frac{p+1}{2}}}I(\epsilon_1,\epsilon)
 \eqe
where $\emin$, $\emax$ are the effective minimum and maximum
energies of the synchrotron photons 
and \eqb \label{Igamma}
 I(\epsilon_1,\epsilon)& = &\int_{\tgammamin}^{\tgammamax}{d\gamma \
\gamma^{-p}\epsilon_1\delta(\epsilon_1-\frac{4}{3}\gamma^2\epsilon)} \nonumber \\
& = &
\frac{1}{2}\left(\frac{\sqrt{3}}{2}\right)^{-p+1}\!\!\!\epsilon_1^{-\frac{p-1}{2}}\epsilon^{\frac{p-1}{2}}
\eqe for $\frac{4}{3}\tgammamin^2 \epsilon < \epsilon_1 <
\frac{4}{3}\tgammamax^2\epsilon$. Strictly speaking, the integral
equals to zero for any other value of $\epsilon_1$. However, if one
uses the complete expression of
 the Inverse Compton emissivity, finds that outside this energy range the intensity is highly reduced but not actually zero. 
 The
lower limit of integration in (\ref{Igamma}) is determined by the
kinematics of the compton scattering and the lower cutoff of the
electron distribution, i.e \eqb \tgammamin=\rm max[\gammamin,
(3\epsilon_1/4\epsilon)^{1/2}]. \eqe The upper limit of integration
is given by \eqb \tgammamax=\rm min[\gammamax,m_{\rm
e}c^2/\epsilon]\eqe which takes into account the effect of the
Klein-Nishina cutoff. For reasons of simplicity we proceed to the
calculation of the SSC spectrum assuming that $\tgammamin=\gammamin$
and $\tgammamax=\gammamax$.

\section{The Compton-synchrotron logarithm}
The integration over the synchrotron photon distribution (see eq.
(\ref{ICtot2}), (\ref{Igamma})) leads to a factor
 \eqb \rm ln \Sigma=
\rm ln\left(\frac{\emax}{\emin}\right) \eqe called the Compton
synchrotron logarithm \citep{gould79}. This quantity takes into
account the effective minimum and maximum energies of synchrotron
photons, which contribute to the principal branch of the SSC
spectrum, i.e to upscattered photons with energies between $\sim
\epsilon_s\gammamin^4$ and $\sim \epsilon_s\gammamax^4$. The actual
extent of the synchrotron power law segment is \eqb \label{esyn}
\epsilon_s \gammamin^2 < \epsilon < \epsilon_s \gammamax^2\eqe where
$\epsilon_s=\frac{eB}{2\pi m_{\rm e}c}$. In the previous section we
showed that only for energies of the upscattered synchrotron photons
between \eqb \label{e1} \frac{4}{3}\gammamin^2 \epsilon < \epsilon_1
< \frac{4}{3}\gammamax^2\epsilon\eqe the SSC spectrum differs
significantly from zero. From eq. (\ref{esyn}), (\ref{e1}) we find
$\epsilon <\epsilon_s\gammamax^2$ and $\epsilon <
3\epsilon_1/4\gammamin^2$.
\begin{enumerate}
\item If $3 \epsilon_1/4\gammamin^2<\epsilon_s\gammamax^2$ then one
sets
 \eqb
 \emax & = & 3 \epsilon_1/4\gammamin^2 \\
\emin & = & \epsilon_s \gammamin^2.
 \eqe
Thus,
 \eqb \Sigma=\frac{3\epsilon_1}{4\epsilon_s\gammamin^4}, \quad
\frac{4}{3}\epsilon_s\gammamin^4 < \epsilon_1<
\frac{4}{3}\epsilon_s\gammamin^2 \gammamax^2 \eqe
\item If $3 \epsilon_1/4\gammamin^2>\epsilon_s\gammamax^2$ then
\eqb \emax=\epsilon_s\gammamax^2\eqe However, synchrotron photons
with energy $\epsilon_s\gammamin^2$ cannot be upscattered to
energies $\epsilon_1> \frac{4}{3}\epsilon_s \gammamin^2
\gammamax^2$. In this case the effective minimum energy of the
synchrotron photons is given by \eqb
\emin=\frac{3\epsilon_1}{4\gammamax^2}\eqe Thus we find
 \eqb
\Sigma=\frac{4\epsilon_s\gammamax^4}{3\epsilon_1}, \quad
\frac{4}{3}\epsilon_s\gammamin^2\gammamax^2 < \epsilon_1<
\frac{4}{3}\epsilon_s\gammamax^4 \eqe
\end{enumerate}
Summarizing,
 \eqb \epsilon_1J_{\rm ic}\! \propto \! \epsilon_1^{-\frac{(p-3)}{2}}\left\{
\begin{array}{ll}\rm
ln\left(\frac{3\epsilon_1}{4\epsilon_s\gammamin^4}\right), &
\frac{4}{3}\epsilon_s\gammamin^4 < \epsilon_1<
\frac{4}{3}\epsilon_s\gammamin^2 \gammamax^2 \\
\phantom{f}&\phantom{d}\\
\phantom{f}&\phantom{d}\\
 \rm ln
\left(\frac{4\epsilon_s\gammamax^4}{3\epsilon_1}\right), &
\frac{4}{3}\epsilon_s\gammamin^2\gammamax^2 < \epsilon_1<
\frac{4}{3}\epsilon_s\gammamax^4
\end{array} \right.
\eqe The function above has a peak at the characteristic energy \eqb
\epsilon_{\rm peak}=\frac{4}{3}\epsilon_s\gammamin^2\gammamax^2\eqe

\bibliographystyle{aa}
\bibliography{grb10c}

\begin{thebibliography}{47}
\expandafter\ifx\csname natexlab\endcsname\relax\def\natexlab#1{#1}\fi

\bibitem[{{Barthelmy} {et~al.}(2005){Barthelmy}, {Cannizzo}, {Gehrels},
  {Cusumano}, {Mangano}, {O'Brien}, {Vaughan}, {Zhang}, {Burrows}, {Campana},
  {Chincarini}, {Goad}, {Kouveliotou}, {Kumar}, {M{\'e}sz{\'a}ros}, {Nousek},
  {Osborne}, {Panaitescu}, {Reeves}, {Sakamoto}, {Tagliaferri}, \&
  {Wijers}}]{Barthelmy05}
{Barthelmy}, S.~D., {Cannizzo}, J.~K., {Gehrels}, N., {et~al.} 2005, \apjl,
  635, L133

\bibitem[{{Blandford} \& {McKee}(1976)}]{blandMckee76}
{Blandford}, R.~D. \& {McKee}, C.~F. 1976, Physics of Fluids, 19, 1130

\bibitem[{{Blumenthal} \& {Gould}(1970)}]{blumgould70}
{Blumenthal}, G.~R. \& {Gould}, R.~J. 1970, Reviews of Modern Physics, 42, 237

\bibitem[{{Butler} \& {Kocevski}(2007)}]{butler2007}
{Butler}, N.~R. \& {Kocevski}, D. 2007, \apj, 668, 400

\bibitem[{{Dermer}(2007)}]{dermer07}
{Dermer}, C.~D. 2007, \apj, 664, 384

\bibitem[{{Dermer} {et~al.}(2000){Dermer}, {B{\"o}ttcher}, \&
  {Chiang}}]{dermbottcher00}
{Dermer}, C.~D., {B{\"o}ttcher}, M., \& {Chiang}, J. 2000, \apj, 537, 255

\bibitem[{{Dermer} \& {Chiang}(1998)}]{dermchiang98}
{Dermer}, C.~D. \& {Chiang}, J. 1998, New Astronomy, 3, 157

\bibitem[{{Eichler} \& {Granot}(2006)}]{eichler06}
{Eichler}, D. \& {Granot}, J. 2006, \apjl, 641, L5

\bibitem[{{Evans} {et~al.}(2009){Evans}, {Beardmore}, {Page}, {Osborne},
  {O'Brien}, {Willingale}, {Starling}, {Burrows}, {Godet}, {Vetere}, {Racusin},
  {Goad}, {Wiersema}, {Angelini}, {Capalbi}, {Chincarini}, {Gehrels}, {Kennea},
  {Margutti}, {Morris}, {Mountford}, {Pagani}, {Perri}, {Romano}, \&
  {Tanvir}}]{Evans09}
{Evans}, P.~A., {Beardmore}, A.~P., {Page}, K.~L., {et~al.} 2009, \mnras, 397,
  1177

\bibitem[{{Evans} {et~al.}(2010){Evans}, {Willingale}, {Osborne}, {O'Brien},
  {Page}, {Markwardt}, {Barthelmy}, {Beardmore}, {Burrows}, {Pagani},
  {Starling}, {Gehrels}, \& {Romano}}]{evans10}
{Evans}, P.~A., {Willingale}, R., {Osborne}, J.~P., {et~al.} 2010, \aap, 519,
  A102+

\bibitem[{{Fan} \& {Piran}(2006)}]{fan_piran06}
{Fan}, Y. \& {Piran}, T. 2006, \mnras, 369, 197

\bibitem[{{Fan} {et~al.}(2008){Fan}, {Piran}, {Narayan}, \& {Wei}}]{fanetal08}
{Fan}, Y., {Piran}, T., {Narayan}, R., \& {Wei}, D. 2008, \mnras, 384, 1483

\bibitem[{{Fenimore} {et~al.}(1993){Fenimore}, {Epstein}, \& {Ho}}]{fenimore93}
{Fenimore}, E.~E., {Epstein}, R.~I., \& {Ho}, C. 1993, \aaps, 97, 59

\bibitem[{{Ghisellini}(2008)}]{ghisellini08}
{Ghisellini}, G. 2008, in American Institute of Physics Conference Series, Vol.
  1000, American Institute of Physics Conference Series, ed. {M.~Galassi,
  D.~Palmer, \& E.~Fenimore}, 448--451

\bibitem[{{Ghisellini} {et~al.}(2007){Ghisellini}, {Ghirlanda}, {Nava}, \&
  {Firmani}}]{ghisellini07}
{Ghisellini}, G., {Ghirlanda}, G., {Nava}, L., \& {Firmani}, C. 2007, \apjl,
  658, L75

\bibitem[{{Giannios} \& {Spitkovsky}(2009)}]{giannios09}
{Giannios}, D. \& {Spitkovsky}, A. 2009, \mnras, 400, 330

\bibitem[{{Gould}(1979)}]{gould79}
{Gould}, R.~J. 1979, \aap, 76, 306

\bibitem[{{Granot} {et~al.}(2006){Granot}, {K{\"o}nigl}, \&
  {Piran}}]{granot_konigl06}
{Granot}, J., {K{\"o}nigl}, A., \& {Piran}, T. 2006, \mnras, 370, 1946

\bibitem[{{Granot} \& {Kumar}(2006)}]{granot_kumar06}
{Granot}, J. \& {Kumar}, P. 2006, \mnras, 366, L13

\bibitem[{{Granot} \& {Sari}(2002)}]{granotsari02}
{Granot}, J. \& {Sari}, R. 2002, \apj, 568, 820

\bibitem[{{Liang} {et~al.}(2007){Liang}, {Zhang}, \& {Zhang}}]{liang07}
{Liang}, E., {Zhang}, B., \& {Zhang}, B. 2007, \apj, 670, 565

\bibitem[{{Lithwick} \& {Sari}(2001)}]{lithwick01}
{Lithwick}, Y. \& {Sari}, R. 2001, \apj, 555, 540

\bibitem[{{Mastichiadis} \& {Kazanas}(2009)}]{mastichiadiskazanas09}
{Mastichiadis}, A. \& {Kazanas}, D. 2009, \apjl, 694, L54

\bibitem[{{Mastichiadis} \& {Kirk}(1995)}]{mastikirk95}
{Mastichiadis}, A. \& {Kirk}, J.~G. 1995, \aap, 295, 613

\bibitem[{{M\'{e}sz\'{a}ros} \& {Rees}(1997)}]{mesz_rees97}
{M\'{e}sz\'{a}ros}, P. \& {Rees}, M.~J. 1997, \apj, 476, 232

\bibitem[{{Nousek} {et~al.}(2006){Nousek}, {Kouveliotou}, {Grupe}, {Page},
  {Granot}, {Ramirez-Ruiz}, {Patel}, {Burrows}, {Mangano}, {Barthelmy},
  {Beardmore}, {Campana}, {Capalbi}, {Chincarini}, {Cusumano}, {Falcone},
  {Gehrels}, {Giommi}, {Goad}, {Godet}, {Hurkett}, {Kennea}, {Moretti},
  {O'Brien}, {Osborne}, {Romano}, {Tagliaferri}, \& {Wells}}]{nousek06}
{Nousek}, J.~A., {Kouveliotou}, C., {Grupe}, D., {et~al.} 2006, \apj, 642, 389

\bibitem[{{Paczynski} \& {Rhoads}(1993)}]{paczynski_rhoads93}
{Paczynski}, B. \& {Rhoads}, J.~E. 1993, \apjl, 418, L5+

\bibitem[{{Panaitescu} \& {Kumar}(2000)}]{panai_kumar00}
{Panaitescu}, A. \& {Kumar}, P. 2000, \apj, 543, 66

\bibitem[{{Panaitescu} \& {M\'{e}sz\'{a}ros}(1998)}]{panai_mesz98}
{Panaitescu}, A. \& {M\'{e}sz\'{a}ros}, P. 1998, \apj, 501, 772

\bibitem[{{Panaitescu} {et~al.}(2006){Panaitescu}, {M{\'e}sz{\'a}ros},
  {Burrows}, {Nousek}, {Gehrels}, {O'Brien}, \& {Willingale}}]{panaietal06}
{Panaitescu}, A., {M{\'e}sz{\'a}ros}, P., {Burrows}, D., {et~al.} 2006, \mnras,
  369, 2059

\bibitem[{{Pe'er} \& {Waxman}(2004)}]{peer04}
{Pe'er}, A. \& {Waxman}, E. 2004, \apj, 613, 448

\bibitem[{{Petropoulou} \& {Mastichiadis}(2009)}]{me09}
{Petropoulou}, M. \& {Mastichiadis}, A. 2009, \aap, 507, 599

\bibitem[{{Piran}(1999)}]{Piran99}
{Piran}, T. 1999, \physrep, 314, 575

\bibitem[{{Piran}(2004)}]{piran05}
{Piran}, T. 2004, Reviews of Modern Physics, 76, 1143

\bibitem[{{Press} {et~al.}(1992){Press}, {Teukolsky}, {Vetterling}, \&
  {Flannery}}]{pressetal}
{Press}, W.~H., {Teukolsky}, S.~A., {Vetterling}, W.~T., \& {Flannery}, B.~P.
  1992, {Numerical recipes in FORTRAN. The art of scientific computing}, ed.
  {Press, W.~H., Teukolsky, S.~A., Vetterling, W.~T., \& Flannery, B.~P. }

\bibitem[{{Rees} \& {M\'{e}sz\'{a}ros}(1992)}]{rees_mesz92}
{Rees}, M.~J. \& {M\'{e}sz\'{a}ros}, P. 1992, \mnras, 258, 41P

\bibitem[{{Sari} \& {Esin}(2001)}]{sariesin01}
{Sari}, R. \& {Esin}, A.~A. 2001, \apj, 548, 787

\bibitem[{{Sari} {et~al.}(1998){Sari}, {Piran}, \& {Narayan}}]{sariPiran98}
{Sari}, R., {Piran}, T., \& {Narayan}, R. 1998, \apjl, 497, L17+

\bibitem[{{Schady} {et~al.}(2007){Schady}, {Mason}, {Page}, {de Pasquale},
  {Morris}, {Romano}, {Roming}, {Immler}, \& {vanden Berk}}]{schady07}
{Schady}, P., {Mason}, K.~O., {Page}, M.~J., {et~al.} 2007, \mnras, 377, 273

\bibitem[{{Shao} {et~al.}(2008){Shao}, {Dai}, \& {Mirabal}}]{shao07}
{Shao}, L., {Dai}, Z.~G., \& {Mirabal}, N. 2008, \apj, 675, 507

\bibitem[{{Sironi} \& {Spitkovsky}(2009)}]{spitkovsky09}
{Sironi}, L. \& {Spitkovsky}, A. 2009, \apjl, 707, L92

\bibitem[{{Spitkovsky}(2008)}]{spitkovsky08}
{Spitkovsky}, A. 2008, \apjl, 682, L5

\bibitem[{{Tagliaferri} {et~al.}(2005){Tagliaferri}, {Goad}, {Chincarini},
  {Moretti}, {Campana}, {Burrows}, {Perri}, {Barthelmy}, {Gehrels}, {Krimm},
  {Sakamoto}, {Kumar}, {M{\'e}sz{\'a}ros}, {Kobayashi}, {Zhang}, {Angelini},
  {Banat}, {Beardmore}, {Capalbi}, {Covino}, {Cusumano}, {Giommi}, {Godet},
  {Hill}, {Kennea}, {Mangano}, {Morris}, {Nousek}, {O'Brien}, {Osborne},
  {Pagani}, {Page}, {Romano}, {Stella}, \& {Wells}}]{tagliaferri05}
{Tagliaferri}, G., {Goad}, M., {Chincarini}, G., {et~al.} 2005, \nat, 436, 985

\bibitem[{{Vaughan} {et~al.}(2006){Vaughan}, {Goad}, {Beardmore}, {O'Brien},
  {Osborne}, {Page}, {Barthelmy}, {Burrows}, {Campana}, {Cannizzo}, {Capalbi},
  {Chincarini}, {Cummings}, {Cusumano}, {Giommi}, {Godet}, {Hill}, {Kobayashi},
  {Kumar}, {La Parola}, {Levan}, {Mangano}, {M{\'e}sz{\'a}ros}, {Moretti},
  {Morris}, {Nousek}, {Pagani}, {Palmer}, {Racusin}, {Romano}, {Tagliaferri},
  {Zhang}, \& {Gehrels}}]{vaughan06}
{Vaughan}, S., {Goad}, M.~R., {Beardmore}, A.~P., {et~al.} 2006, \apj, 638, 920

\bibitem[{{Wijers} \& {Galama}(1999)}]{wijers_galama99}
{Wijers}, R.~A.~M.~J. \& {Galama}, T.~J. 1999, \apj, 523, 177

\bibitem[{{Zhang} {et~al.}(2006){Zhang}, {Fan}, {Dyks}, {Kobayashi},
  {M{\'e}sz{\'a}ros}, {Burrows}, {Nousek}, \& {Gehrels}}]{zhang06}
{Zhang}, B., {Fan}, Y.~Z., {Dyks}, J., {et~al.} 2006, \apj, 642, 354

\bibitem[{{Zhang} {et~al.}(2007){Zhang}, {Liang}, \& {Zhang}}]{zhangliang07}
{Zhang}, B., {Liang}, E., \& {Zhang}, B. 2007, \apj, 666, 1002

\end{thebibliography}
\end{document}